\documentclass[12pt]{article} 
\usepackage{amsmath}
\usepackage{amsfonts}
\usepackage{dsfont}
\usepackage{cite}
\usepackage{epsfig}
\usepackage{epic}
\usepackage{eepic}
\usepackage{axodraw}
\usepackage{subfigure}
\usepackage{rotating}

\def\be{\begin{equation}}
\def\ee{\end{equation}}
\newcommand{\bea}{\begin{eqnarray}}
\newcommand{\eea}{\end{eqnarray}}
\newcommand{\col}{~,}
\newcommand{\pnt}{~.}
\newcommand{\AdS}{\text{AdS}}
\newcommand{\AdSS}{\AdS_{d+1}\times \text{S}^{d'+1}}
\newcommand{\twob}{\text{II}\,\text{B}}

\newcommand{\preparderiv}[1]{\frac{\partial}{\partial #1}}
\newcommand{\doublepreparderiv}[1]{\frac{\partial^2}{\partial {#1}^2}}
\newcommand{\hypergeometric}[4]{F\big(#1,#2;#3;#4\big)}
\newlength{\diameter}
    
\newlength{\neglength}
         
\makeatletter
\newcommand{\redefinelabel}[1]{
  \def\@currentlabel{#1}}
\makeatother

\DeclareMathOperator{\sgn}{sgn}
\newcommand{\de}{\operatorname{d}\!}
\begin{document}
\begin{titlepage}
\begin{flushright}
HU Berlin-EP-03/41\\ 
\end{flushright}
\mbox{ }  \hfill hep-th/0307229
\vspace{5ex}
\Large
\begin {center}     
{\bf On the propagator of a scalar field in $\AdS\times\text{S}$ and in its
  plane wave limit}
\end {center}
\large
\vspace{1ex}
\begin{center}
Harald Dorn, Mario Salizzoni and Christoph Sieg \footnote{dorn@physik.hu-berlin.de, sali@physik.hu-berlin.de, csieg@physik.hu-berlin.de}
\end{center}
\begin{center}
Humboldt--Universit\"at zu Berlin, Institut f\"ur Physik\\
Newtonstra\ss e 15, D-12489 Berlin\\[2mm]  
\end{center}
\vspace{4ex}
\rm
\begin{center}
{\bf Abstract}
\end{center} 
\normalsize 
We discuss the scalar propagator on generic $\AdSS$ backgrounds.
For the conformally flat situations and masses corresponding to Weyl
invariant actions, the propagator is powerlike in the sum of the
chordal distances with respect to $\AdS_{d+1}$ and $\text{S}^{d'+1}$.
In these cases we analyze its source structure.
In all other cases the propagator depends on both chordal distances
separately. There an explicit formula is found for certain
special mass values. For pure AdS we show how the well known
propagators in the Weyl invariant case can be expressed as linear
combinations of simple powers of the chordal distance. For
$\text{AdS}_5\times\text{S}^5$ we relate our propagator to the expression
 in the plane wave limit and find a geometric interpretation of the variables
occurring in the known explicit construction on the plane wave.
As a byproduct of comparing different techniques, including the
KK mode summation,  a theorem for summing certain products of
Legendre and Gegenbauer functions is derived.
\vfill
\end{titlepage} 
\section{Introduction}
The AdS/CFT correspondence \cite{Maldacena:1998re} relates $\mathcal{N}=4$
super Yang-Mills gauge theory in Min\-kow\-ski space to type  $\twob$ string
theory in $\AdS_5\times\text{S}^5$ with some RR background flux. In the
supergravity approximation one handles the fields in a Kaluza-Klein mode
expansion with respect to the $\text{S}^5$. For calculations on the
supergravity side the propagators of the whole spectrum of fields in the
$\AdS_5$  background are an essential technical ingredient. The simplest case
to start with is of course the well known scalar propagator
\cite{Burgess:1985ti,D'Hoker:2002aw}. Explicit tests of the AdS/CFT
correspondence, beyond the supergravity approximation, remain a difficult
task, since the relevant string spectrum in general is not available. 
In a limit of this correspondence, proposed by Berenstein, Maldacena and Nastase (BMN limit)
\cite{Berenstein:2002jq}, the $\AdS_5\times\text{S}^5$ background itself is
transformed via a Penrose limit \cite{Penrose:1976} to a certain plane wave
background found by Blau, Figueroa-O'Farrill, Hull and Papadopoulos
\cite{Blau:2001ne,Blau:2002dy,Blau:2002mw}. In this background, for brevity 
called `the plane wave',
string theory is exactly quantizable \cite{Metsaev:2001bj}, and thus enables
independent checks of the duality, including string effects. In this plane
wave background the separation between the $\AdS_5$ and the $\text{S}^5$ part
breaks down, and one has to take the limit on full 10-dimensional
$\AdS_5\times\text{S}^5$ objects.  

One of the crucial unsolved questions in this setting concerns the issue
of holography 
\cite{Berenstein:2002sa,Das:2002cw,Leigh:2002pt,Kiritsis:2002kz,Dobashi:2002ar}.
In the Penrose limiting process the old 4-dimensional conformal boundary is
put beyond the new plane wave space, which by itself has a one-dimensional
conformal boundary. In \cite{Dorn:2003ct} we started to investigate this issue
and found, for each point remaining in the final plane wave, a degeneration of
the cone of boundary reaching null geodesics into a single direction. To
continue this program beyond geometric properties, we now want to study the
limiting process  for field theoretical propagators. 

The so called bulk-to-boundary propagator plays an essential role
in the holographic description of the AdS/CFT correspondence and is therefore
a quantity of particular interest. However, as follows from the results 
in \cite{Dorn:2003ct}, a reasonable Penrose limit cannot be taken due to the 
fact that one of its legs ends at the boundary and hence lies outside the 
region of convergence to the plane wave. The situation is different for the 
bulk-to-bulk propagator. One has the choice to let both legs end within the
region that converges to the plane wave in the limit. Although this propagator
first seems to be of minor importance for the realization of holography and
even for the computation of correlation functions \cite{D'Hoker:1999ni}, it
might be very useful  
for defining a bulk-to-boundary propagator in the plane wave. A hint that this
could be a promising direction is, that in pure $\AdS_{d+1}$ the knowledge 
of the bulk-to-bulk propagator is sufficient for deriving the corresponding 
bulk-to-boundary propagator. 

Due to the existence of such a relation it seems to be worthwhile 
to study also in $\AdS\times\text{S}$ the 
the bulk-to-bulk propagator to get information about the
bulk-boundary correspondence in the plane wave limit. For brevity when we
talk about `the propagator' in the following we always understand it as
the bulk-to-bulk one.

The scalar propagator in
the  plane wave has been constructed in \cite{Mathur:2002ry} by a direct
approach leaving the issue of its derivation via a limiting process from
$\AdS_5\times\text{S}^5$ as an open problem.  

Motivated by the above given questions, and because it is an 
interesting problem in its 
own right, we will study in this paper the construction of the scalar
propagator on $\AdSS$ spaces with radii $R_1$ and $R_2$, respectively.
Allowing for generic dimensions $d$ and $d'$ as well as generic curvature
radii  $R_1$ and $R_2$ is very helpful to understand the general mechanism for
the construction of the propagator. Of course only some of these spacetimes
are parts of consistent supergravity backgrounds. 

In Section \ref{waveeqapp} we will focus on the differential equation defining
the scalar propagator. Within this Section we will be able to find the
propagator in conformally flat situations, i.e. $R_1=R_2$ and for masses
corresponding to Weyl invariant actions.

The next two Sections mainly serve as a kind of 
interpretation of the results
of Section \ref{waveeqapp}. Using Weyl invariance, we map patchwise to 
flat space in Section \ref{weylana} and globally to the Einstein Static
Universe (ESU) in Section \ref{relESU}. This includes a discussion of  
global aspects of the solutions, like their boundary conditions and 
$\delta$-source structure.  

With the hope to get the propagator for generic masses, in Section
\ref{modesummation} we study its KK mode sum. We will be able to perform the
sum for a linear relation between the conformal dimension of the KK mode
and the quantum number parameterizing the eigenvalue of the Laplacian on
the sphere. Beyond the cases treated in the previous Sections
this applies to certain additional mass values, but fails to solve
the full generic problem. As a byproduct, the comparison with
the result of Section \ref{waveeqapp} yields a theorem on the summation of
certain products of Gegenbauer and Legendre functions.  

In Section \ref{pwlimit} we will discuss the plane wave limit of 
$\AdS_5\times\text{S}^5$ in brief. We will explicitly show that the massless
propagator on the full spacetime indeed reduces to the expression of
\cite{Mathur:2002ry}. Furthermore, we will present the limit of the full
differential equation which is fulfilled by the propagator of massive scalar
fields given in \cite{Mathur:2002ry}. 
Finally, our results will be summarized in Section \ref{concl}. 

In Appendix \ref{bulkboundproprel} a detailed derivation of the relation
between the bulk-to-bulk and the bulk-to-boundary propagator in the case 
of pure Euclidean $\AdS_{d+1}$ is presented.
Appendix \ref{userel} summarizes some of the
relevant formulae for hypergeometric functions and spherical harmonics we
needed for the analysis. Furthermore we will sketch an independent proof of
the theorem that has been extracted from Sections \ref{waveeqapp} and
\ref{modesummation}. Appendix \ref{chordaldistpw} contains a short review
of the embedding of the plane wave in flat spacetime.

%%%%%%%%%%%%%%%%%%%%%%%%%%%%%%%%%%%%%%%%%%%
%%%%%%%%%%%%%%%%%%%%%%%%%%%%%%%%%%%%%%%%%%%
\section{The differential equation for the propagator and its solution}
\label{waveeqapp}
\subsection{The scalar propagator on $\AdSS$}
\label{AdSSderiv}
The scalar propagator is defined as the solution 
of\footnote{This normalization is 
consistent with the definition in \cite{D'Hoker:2002aw}, because in a
continuation to Euclidean space the factor $i$ in front of the $\delta$-source 
becomes $-1$ as in \eqref{Gfuncdiffeq}.}
\begin{equation}\label{waveeq}
(\Box_z-M^2)G(z,z')=\frac{i}{\sqrt{-g}}\delta(z,z')\col
\end{equation}
with suitable boundary conditions at infinity. $\Box_z$ denotes the d'Alembert
operator on $\AdSS$, acting on the first argument of the propagator $G(z,z')$.
In the following we denote the coordinates referring to the $\AdS_{d+1}$
factor by $x$ and those referring to the $\text{S}^{d'+1}$  factor by $y$,
i.e. $z=(x,y)$. We first look for solutions at $z\neq z'$ and discuss the
behaviour at $z=z'$ afterwards. 

$\AdS_{d+1}$ and $\text{S}^{d'+1}$ can be interpreted as embeddings
respectively in $\mathds{R}^{2,d}$ and in $\mathds{R}^{d'+2}$ with the help of
the constraints\footnote{More precisely, $\AdS_{d+1}$ is the universal
  covering of the hyperboloid in $\mathds{R}^{2,d}$.}  
\begin{equation}\label{defeq}
-X_0^2-X_{d+1}^2+\sum_{i=1}^d
 X_i^2=-R_1^2\col\qquad\sum_{i=1}^{d'+2}Y_i^2=R_2^2\col 
\end{equation}
where $X=X(x)$, $Y=Y(y)$ depend on the coordinates $x$ and $y$, respectively.
We define the chordal distances on both spaces to be 
\begin{equation}\label{uv}
u(x,x')=(X(x)-X(x'))^2\col\qquad v(y,y')=(Y(y)-Y(y'))^2\pnt
\end{equation}
The distances have to be computed with the corresponding flat metrics of the
embedding spaces that can be read off from  \eqref{defeq}. 
The chordal distance $u$ is a unique function of $x$ and $x'$ if one restricts
oneself to the hyperboloid. On the 
universal covering it is continued as a periodic function.
For later use we note that on the hyperboloid and on the sphere the 
antipodal points $\tilde x$ and $\tilde y$ to given points $x$ and $y$ are
defined by changing the sign of the embedding coordinates $X$ and $Y$
respectively. 
From \eqref{uv} one then finds with
$\tilde u=u(x,\tilde x')$, 
$\tilde v=v(y,\tilde y')$
\begin{equation}\label{uvaprel}
%\begin{aligned}
u+\tilde u=-4R_1^2\col\qquad v+\tilde v=4R_2^2\pnt
%\end{aligned}
\end{equation}
Using the homogeneity and isotropy of both $\AdS_{d+1}$ and $\text{S}^{d'+1}$
it is clear that the propagator can depend on $z,z'$ only via the chordal
distances $u(x,x')$ and $v(y,y')$. Strictly speaking this at first applies
only if $\AdS_{d+1}$ is restricted to the hyperboloid. Up to subtleties due 
to time ordering (see the end of Section \ref{relESU}) 
this remains true also on the universal covering.
The d'Alembert operator then simplifies to
\begin{equation}\label{dalemchordal}
\begin{aligned}
\Box_z&=\Box_x+\Box_y\col\\
\Box_x&=2(d+1)\Big(1+\frac{u}{2R_1^2}\Big)\preparderiv{u}+\Big(\frac{u^2}{R_1^2}+4u\Big)\doublepreparderiv{u}\col\\
\Box_y&=2(d'+1)\Big(1-\frac{v}{2R_2^2}\Big)\preparderiv{v}-\Big(\frac{v^2}{R_2^2}-4v\Big)\doublepreparderiv{v}\pnt
\end{aligned}
\end{equation}

One can now ask for a solution of \eqref{waveeq} at $z\neq z'$ that only
depends on the total chordal distance $u+v$. Indeed, using
\eqref{dalemchordal}, it is easy to derive that such a solution exists if and
only if
\begin{equation}\label{conditions}
R_1=R_2=R\col\qquad M^2=\frac{d'^2-d^2}{4R^2}\pnt
\end{equation}
Furthermore, it is necessarily powerlike and given by
\begin{equation}
G(z,z')\propto (u+v)^{-\frac{d+d'}{2}}\pnt
\end{equation}
Extending this to $z=z'$ we find just the right power for the short
distance singularity to generate the $\delta$-function on the r.h.s.\ of 
\eqref{waveeq}. Hence after fixing the normalization we end up with
\begin{equation}\label{AdSSprop}
G(z,z')=\frac{\Gamma(\frac{d+d'}{2})}{4\pi^{\frac{d+d'}{2}+1}}\frac{1}{(u+v+i\varepsilon(t,t'))^\frac{d+d'}{2}}\pnt
\end{equation}
Note that due to \eqref{uvaprel} besides the singularity at $z=z'$ there is 
another one at the total antipodal point where 
$z=\tilde z'=(\tilde x', \tilde y')$. 
We have introduced an $i\varepsilon$-prescription by replacing $u\to
u+i\varepsilon$, where $\varepsilon$ depends explicitly on time. 
We will comment on this in Section \ref{relESU}. In particular, we will see
that on the universal covering of the hyperboloid the singularity at the 
total antipodal point does not lead to 
an additional $\delta$-source on the r.h.s.\ of \eqref{waveeq}.

Scalar fields with mass $m^2$ in $\AdS_{d+1}$ via the $\AdS/\text{CFT}$ correspondence
are related to CFT fields with conformal dimension
\begin{equation}\label{conformaldim}
\Delta_\pm(d,m^2)=\frac{1}{2}\Big(d\pm\sqrt{d^2+4m^2R_1^2}\Big)\pnt
\end{equation}
Note that the exponent of $(u+v)$ in the denominator of the propagator
\eqref{AdSSprop} 
is just equal to $\Delta_+(d,M^2)$. From the $\AdS_{d+1}$ point of view
the $(d+d'+2)$-dimensional mass $M^2$ is the mass of the KK zero mode
of the sphere. We will say more on these issues in Section 
\ref{modesummation}.\\ 

For completeness let us add another observation. Disregarding for a moment the 
source structure, 
under the conditions \eqref{conditions} there is a solution of \eqref{waveeq},
that depends only on $(u-v)$. 
The explicit form is
\begin{equation}\label{AdSSmirprop}
\tilde G(z,z')\propto\frac{1}{(u-v+4R^2+i\varepsilon(t,t'))^\frac{d+d'}{2}}\pnt
\end{equation}
It has the same asymptotic falloff as \eqref{AdSSprop}. But due to
\eqref{uvaprel} it has singularities only at the semi-antipodal points
where $z=z'_\text{s}=(x',\tilde y')$ and 
$z=\tilde z'_\text{s}=(\tilde x',y')$. 
We will say more on $\tilde G(z,z')$ in Sections \ref{weylana} and
\ref{relESU}. \\

At the end of this Subsection we give a simple interpretation of the
conditions \eqref{conditions}. The equality of the radii is exactly the
condition for conformal flatness of the complete product space $\AdSS$ as a
whole. Describing the AdS metric in Poincar\'e coordinates
($x=(x^0,x^1,\dots,x^{d-1},x_{\perp })$) one finds 
\begin{equation}\label{AdSmetric}
\de s^2=\frac{R_1^2}{x_{\perp}^2}\big(-(\de x^0)^2+\de x_{\perp}^2+\de\vec x^2\big)\pnt
\end{equation}
For $\AdSS$ one thus obtains
\begin{equation}\label{AdSSmetric}
\de s^2=\frac{R_1^2}{x_{\perp}^2}\left ({-(\de x^0)^2+\de x_{\perp}^2+\de\vec
    x^2}+\frac{R_2^2}{R_1^2}x_{\perp}^2\de\Omega_{d'+1}^2\right )\col 
\end{equation}
which is obviously conformally flat if $R_1=R_2$. That this is also necessary
for conformal flatness follows from an analysis of the corresponding Weyl
tensor. 
Furthermore, the mass condition just singles out the case of a scalar
field coupled in Weyl invariant manner to the gravitational background.
The corresponding $D$-dimensional action is 
\begin{equation}
S=-\frac{1}{2}\int\de^Dz\,\sqrt{-g}\Big[g^{\mu\nu}\partial_\mu\phi\partial_\nu\phi+\frac{D-2}{4(D-1)}\mathcal{R}\phi^2\Big]\pnt
\end{equation}
Inserting the constant curvature scalar $\mathcal{R}$ for $\AdSS$
with equal radii one gets for the mass just the value in 
\eqref{conditions}.

Altogether in this Subsection we have constructed the scalar $\AdSS$
propagator for the case of Weyl invariant coupling to the metric 
in conformally flat situations. The Weyl invariant coupled field is the 
natural generalization of the massless field in flat space. 

\subsection{A remark on the propagator on pure $\AdS_{d+1}$}
\label{AdSderiv}
Having found for $\AdSS$ such a simple expression for
the scalar propagator, one is wondering whether the well known AdS propagators
can also be related to simple powers of the chordal distance.

The general massive scalar propagator on pure
$\AdS_{d+1}$ space corresponding to the two distinct conformal dimensions
$\Delta_{\pm}$ defined in \eqref{conformaldim} with generic mass
values is given by \cite{Burgess:1985ti,D'Hoker:2002aw}
\begin{equation}\label{AdSprop}
G_{\Delta_\pm}(x,x')=\frac{\Gamma(\Delta_\pm)}{R_1^{d-1}2\pi^\frac{d}{2}\Gamma(\Delta_\pm-\frac{d}{2}+1)}\Big(\frac{\xi}{2}\Big)^{\Delta_\pm}\hypergeometric{\tfrac{\Delta_\pm}{2}}{\tfrac{\Delta_\pm}{2}+\tfrac{1}{2}}{\Delta_\pm-\tfrac{d}{2}+1}{\xi^2}\col\quad\xi=\frac{2R_1^2}{u+2R_1^2}\pnt 
\end{equation}
 
Again, here a powerlike solution of \eqref{waveeq} (but now using only the AdS
d'Alembert operator of \eqref{dalemchordal} and replacing $M^2$ by the AdS 
mass $m^2$) 
exists for the Weyl invariant coupled mass 
value\footnote{Pure AdS spaces are conformally flat.}
\begin{equation}\label{AdSWeylmass}
m^2=\frac{1-d^2}{4R_1^2}\pnt
\end{equation}
The related value for the conformal dimension from \eqref{conformaldim}
is then $\Delta_{\pm}=\frac{d\pm 1}{2}$. The powerlike solution is given by
\begin{equation}\label{AdSprop2}
G(x,x')=\frac{\Gamma(\frac{d-1}{2})}{4\pi^{\frac{d+1}{2}}}\frac{1}{(u+i\varepsilon(t,t'))^\frac{d-1}{2}}\pnt
\end{equation}%
In contrast to the $\AdSS$ case here the exponent of $u$ is given by
$\Delta_-(d,m^2)$. We have again kept the option of a time dependent
$i\varepsilon(t,t')$ and will comment on it in Section \ref{relESU}.

The above solution can indeed be obtained from \eqref{AdSprop} by taking the
sum of the expressions for $\Delta_+$ and $\Delta_-$. 
In addition one finds another simple structure by taking the difference. 
They are given by 
\begin{equation}\label{AdSsuperpos}
\begin{aligned}
\frac{1}{2}(G_{\Delta_-}+G_{\Delta_+})&=\frac{\Gamma(\frac{d-1}{2})}{4\pi^{\frac{d+1}{2}}}\frac{1}{(u+i\varepsilon(t,t'))^\frac{d-1}{2}}\col\\
\frac{1}{2}(G_{\Delta_-}-G_{\Delta_+})&=\frac{\Gamma(\frac{d-1}{2})}{4\pi^{\frac{d+1}{2}}}\frac{1}{(u+4R^2+i\varepsilon(t,t'))^\frac{d-1}{2}}\pnt
\end{aligned}
\end{equation}
Both expressions are derived by using \eqref{AdSprop} and 
\eqref{hypergeom1} of Appendix 
\ref{userel}. 
The first combination has the right short distance singularity to be a solution
of \eqref{waveeq}. The second combination resembles \eqref{AdSSmirprop}.
We will say more on this in Sections \ref{weylana} and \ref{relESU}. 

\subsection{Comment on masses and conformal dimensions on $\AdS_{d+1}$}
\label{BFbounds}
On AdS spaces one has to respect the Breitenlohner-Freedman 
bounds \cite{Breitenlohner:1982bm,Breitenlohner:1982jf}.
To get real values for $\Delta_\pm$ requires
\begin{equation}\label{BFbound}
m^2\geq -\frac{d^2}{4R_1^2}\pnt
\end{equation}
Furthermore, the so called unitarity bound requires
\begin{equation}\label{splitbound}
\Delta >\frac{d-2}{2}\pnt
\end{equation}
This implies that for $-\frac{d^2}{4R_1^2}\leq m^2 < \frac{4-d^2}{4R_1^2}$
both, $\Delta _+$ and $\Delta _-$ are allowed. On the other side for $\frac
{4-d^2}{4R_1^2}\leq m^2$ only $\Delta _+$ is allowed.

The masses for Weyl invariant coupling are $\frac{1-d^2}{4R_1^2}$ and
$\frac{d'^2-d^2}{4R_1^2}$ for $\AdS_{d+1}$ and $\AdSS$,
respectively. Hence in our Weyl invariant cases for pure AdS 
$\Delta _+$ and $\Delta _-$ are allowed while for $\AdSS$ with $d'>1$
only $\Delta _+$ is allowed. 

%%%%%%%%%%%%%%%%%%%%%%%%%%%%%%%%%%%%%%%%%%%%%    
%%%%%%%%%%%%%%%%%%%%%%%%%%%%%%%%%%%%%%%%%%%
\section{Derivation of the propagator from the flat space one}
\label{weylana}
In the previous Section we have shown that a simple powerlike solution of
\eqref{waveeq} can be found if the underlying spacetime is $\AdS_{d+1}$ or a
conformally flat product space $\AdSS$  and if the corresponding scalar field 
is Weyl invariant coupled to the curvature of the background. Both properties
allow for a mapping of the differential equation, the scalar field and the
propagator to flat space. The other way around, one can use Weyl invariance in
this special case to construct the propagator of Weyl invariant coupled fields
on conformally flat backgrounds from the flat space massless propagator.  

We will use this standard construction to rederive the $\AdSS$ expressions
 \eqref{AdSSprop} and \eqref{AdSSmirprop} from the flat space solutions.

The relevant Weyl transformation in a $D$-dimensional manifold is
\begin{equation}\label{weyltraf1}
g_{\mu\nu}\to \varrho\,g_{\mu\nu}\col\qquad
\phi\to\phi'=\varrho^\frac{2-D}{4}\phi\pnt 
\end{equation}
If then the metric is of the form $g_{\mu\nu}(z)=\varrho(z)\,\eta_{\mu\nu}$ one
finds the following relation between the propagator in curved and flat space
\begin{equation}
G(z,z')=\big(\varrho(z)\,\varrho(z')\big)^{\frac{2-D}{4}}\,G_\text{flat}(z,z')\col\qquad
G_\text{flat}(z,z')=\frac{\Gamma(\frac{D-2}{2})}{4\pi^{\frac{D}{2}}}\frac{1}{((z-z')^2+i\epsilon)^{\frac{D-2}{2}}}
\pnt
\end{equation}
It can be derived either by formal manipulations with the corresponding
functional integral or by using the covariance properties of the defining
differential equation.\footnote{Of course, the discussion has to be completed
  by considering also the boundary conditions.}

Applying the formula first to pure AdS one gets in Poincar\'e coordinates
\eqref{AdSmetric}
\begin{equation}
G(x,x')=\frac{\Gamma(\frac{d-1}{2})}{R_1^{d-1}4\pi^{\frac{d+1}{2}}}
\Big(\frac{1}{x_\perp x'_\perp}\big[(x_\perp-x'_\perp)^2-(x^0-x'^0)^2+(\vec
x-\vec x')^2+i\epsilon\big]\Big)^{\frac{1-d}{2}}\pnt
%\left(\frac{x_\perp}{x'_\perp}+\frac{x'_\perp}{x_\perp}-2+\frac{(\vec x-\vec
%    x')^2-(x^0-x'^0)^2}{x_\perp
%    x'_\perp}+i\epsilon\right)^{\frac{1-d}{2}}\pnt 
\end{equation}
Using the relation between Poincar\'e coordinates and the coordinates in the
embedding space, see e.g.\ \cite{Aharony:1999ti}, it is straightforward to
verify that this with \eqref{uv} is equal to \eqref{AdSprop2}.\\ 

The Poincar\'e patch of pure $\AdS_{d+1}$
is conformal to a flat half space with $x_\perp\ge0$.
$x_\perp=0$ corresponds to the conformal boundary of AdS.
Let us first disregard that the flat half space represents only one half
of $\AdS_{d+1}$ and discuss global issues later. 
We can then implement either Dirichlet or Neumann boundary
conditions by the standard mirror charge method. 
To $x=(x_\perp,x^0,x^1,\dots,x^{d-1})$ we relate the mirror 
point\footnote{Using $x_\perp<0$ for parameterizing the second Poincar\'e 
patch the mirror point is at the antipodal position on the hyperboloid.}
\begin{equation}\label{mirror}
\tilde x=(-x_\perp,x^0,x^1,\dots,x^{d-1})
\end{equation}
and the mirror propagator by
\begin{equation}
\tilde G_\text{flat}(x,x')=G_\text{flat}(x,\tilde x')\pnt
\end{equation}
Then $\frac{1}{2}(G_{\Delta_-}-G_{\Delta_+})$ in the second line of
\eqref{AdSsuperpos} turns out to be just the Weyl transformed version of
$\tilde G_\text{flat}(x,x')$. Equivalently we can state, that
$G_{\Delta_+}$ and $G_{\Delta_-}$ are the Weyl transformed versions
respectively of the
Dirichlet and Neumann propagator in the flat halfspace.

The situation is different for $\AdSS$ spacetimes. According to
\eqref{AdSSmetric}, $x_{\perp}\ge0$ becomes a
radial coordinate of a full $(d'+2)$-dimensional flat subspace of a total 
space with coordinates 
\begin{equation}\label{zcoord}
z=\big(x_0,\vec x,x_\perp\tfrac{\vec Y}{R}\big)\col
\end{equation}
where $\vec Y^2=R^2$
are the embedding coordinates of $S^{d'+1}$. The boundary of
the  AdS part is mapped to the origin of the $(d'+2)$-dimensional subspace. 
Similarly to the pure AdS case, $G(z,z')$ from
\eqref{AdSSprop} is the Weyl transform of $G_\text{flat}(z,z')$. 
To see this one has to cast the length square on the $(d'+2)$-dimensional
subspace, which appears in the denominator of the propagator, into the form
\begin{equation}
\frac{1}{R^2}(x_\perp\vec Y-x'_\perp\vec
  Y')^2=x_\perp^2+x'^2_\perp-2\frac{x_\perp x'_\perp}{R^2}\vec Y\vec Y'
=(x_\perp-x'_\perp)^2+\frac{x_\perp x'_\perp}{R^2}v\pnt
\end{equation} 
%where we have used \eqref{uv}.
In addition, with 
\begin{equation}\label{twistapzcoord}
z_\text{s}=
\big(x_0,\vec x,-x_\perp\tfrac{\vec Y}{R}\big)\col\qquad\tilde
G_\text{flat}(z,z')=G_\text{flat}(z,z'_\text{s})
\end{equation}
we find that the second simple solution \eqref{AdSSmirprop} is the Weyl
transformed version of $\tilde G_\text{flat}(z,z')$.

The coordinates \eqref{zcoord} and
\eqref{twistapzcoord} are related by replacing $\vec Y$ by $-\vec Y$, i.e.
$z_\text{s}$ is related to $z$ by going to the antipodal point 
in the sphere, according to the definition of $z_\text{s}$ after 
\eqref{AdSSmirprop}. 
The two points $z$, $z_\text{s}$ are elements of $\mathds{R}^{d+d'+2}$ 
lying in the first Poincar\'e patch where $x_\perp\ge0$.

As we mentioned before, one has to be careful with global issues. We work in
the Poincar\'e patch that only covers points with $x_\perp\ge0$.
It is easy to see that the coordinates \eqref{zcoord} of $z$ and
\eqref{twistapzcoord} of $z_\text{s}$ 
remain unchanged if one simultaneously replaces $x_\perp$
by $-x_\perp$ and $\vec Y$ by $-\vec Y$. 
This operation switches from $z$ and $z_\text{s}$ respectively to the total
antipodal positions $\tilde z$ and $\tilde z_\text{s}$, that
are covered by a second Poincar\'e patch with $x_\perp<0$. 
Thus the latter points, being elements of the complete manifold,
are not covered by the first Poincar\'e patch.
%To a given point $z$ one thus finds the three other 
%special points that are situated at the antipodal positions in $\AdS_{d+1}$
%and $\text{S}^{d'+1}$. 
%The total antipodal point $\tilde z$ is given by
%switching simultaneously to the antipodal point within both factors.
%$z_\text{s}$ and its total antipodal point $\tilde z_\text{s}$ 
%are reached from $z$ by switching to the antipodal position respectively 
%within $\text{S}^{d'+1}$ and within $\AdS_{d+1}$. We will call $z_\text{s}$
%and $\tilde z_\text{s}$ the semi-antipodal points w.r.t.\ the point $z$. 
In the context of pure $\AdS_{d+1}$, the mirror point $\tilde x$ in
\eqref{mirror} related to $x$ is outside of the first Poincar\'e patch 
but it is still a point in $\AdS_{d+1}$ covered by the second Poincar\'e patch.
Hence $\tilde x$ is not an element of the flat half space that is 
conformal to the first Poincar\'e patch.
We will now analyze the global issues more carefully by
working with the corresponding ESU.

\section{Relation to the ESU}
\label{relESU}
%As discussed in Subsections \ref{AdSspacetime} and \ref{AdSSspacetime},
$\AdS_{d+1}$ and $\AdSS$ with $R_1=R_2$ 
are conformal to respectively one half and to the 
full ESU of the corresponding dimension.
This can be easily seen in a 
certain set of global coordinates where the metric of $\AdS_{d+1}$ assumes
the form
\begin{equation}\label{AdSmetricglobalcoord}
%\de s_\AdS^2&=R_1^2\big(-\cosh^2\rho\de
%t^2+\de\rho^2+\sinh^2\rho\de\Omega_{d-1}^2\big)\col 
\de s_\AdS^2=R_1^2\sec^2\bar\rho\big(-\de
t^2+\de\bar\rho^2+\sin^2\bar\rho\de\Omega_{d-1}^2\big)\col 
\end{equation}
where $0\le\bar\rho<\frac{\pi}{2}$. The corresponding ESU has the topology 
$\mathds{R}\times\text{S}^d$ and its metric is 
given by the expression in parentheses. 
The conformal map between $\AdS_{d+1}$ and $\text{ESU}_{d+1}$ 
has been used in \cite{Avis:1978yn} at $d=3$ to find consistent
quantization schemes on $\AdS_4$. In case of the Weyl invariant mass value
\eqref{AdSWeylmass} the quantization prescription
on the ESU leads to two different descriptions for pure AdS. One can either 
choose transparent boundary conditions or reflective boundary conditions 
at the image of the AdS boundary. The reflectivity
of the boundary is guaranteed for either Dirichlet or Neumann boundary
conditions. This is realized by choosing a subset of modes with definite 
symmetry properties, whereas in the transparent case all modes are used. 
Quantization in the reflective case leads one to the solutions 
$G_{\Delta_\pm}$.  
These results motivate why we will work on the ESU in the following. 
We will find the antipodal points and see how the mirror charge construction 
works. Then we will discuss what this implies for the well known propagators 
in $\AdS_{d+1}$ and our solutions for $\AdSS$ in the Weyl invariant cases.
In the coordinates \eqref{AdSmetricglobalcoord} a point $\tilde x$ 
antipodal to the point $x=(t,\bar\rho,x_\Omega)$ in $\AdS_{d+1}$ is given by
\begin{equation}\label{AdSantipodglobalcoord}
\tilde x=(t+\pi,\bar\rho,\tilde x_\Omega)\col  
\end{equation} 
where $x_\Omega$ denotes the angles of the $(d-1)$-dimensional subsphere of 
$\AdS_{d+1}$ with embedding coordinates $\omega_i$, 
%(see\eqref{Xinglobalcoord})
which then obey
%such that one finds\footnote{See \eqref{Santipodal} for an explicit relation
%  between the angles.}
\begin{equation}\label{AdSsubsphereantipodalrel}
\omega_i(\tilde x_\Omega)=-\omega_i(x_\Omega)\pnt
\end{equation}
The above relation \eqref{AdSantipodglobalcoord} must not be confused with the
relation between two points  
that are antipodal to each other on the sphere of the ESU at fixed time. 

\begin{figure}
\begin{center}
\subfigure[]{\label{fig:AdSESU}%
\begin{picture}(200,180)(0,0)\scriptsize
\SetWidth{1}
%\Boxc(100,90)(200,180)
\SetOffset(0,0)
\SetScale{1}
\SetWidth{0}
\GBox(70,20)(130,160){0.8}
\SetWidth{1}
%\Line(40,25)(160,25)
\Text(40,10)[]{$\pi$}
\Text(70,10)[]{$\frac{\pi}{2}$}
\Text(100,10)[]{$\bar\rho=0$}
\Text(130,10)[]{$\frac{\pi}{2}$}
\Text(160,10)[]{$\pi$}
\Text(70,165)[]{$\omega<0$}
%\Text(55,165)[]{$Y<0$}
\Text(130,165)[]{$\omega>0$}
%\Text(100,165)[]{$Y>0$}
%\Text(145,165)[]{$Y<0$}
\Line(40,20)(40,160)
\Line(70,20)(70,160)
\Line(100,20)(100,160)
\Line(130,20)(130,160)
\Line(160,20)(160,160)
\DashLine(40,35)(80,35){4}\Text(30,35)[cr]{$t$}
\DashLine(120,35)(160,35){4}
\DashLine(40,95)(160,95){4}\Text(30,95)[cr]{$t+\pi$}
\DashLine(120,35)(80,35){1}
\DashLine(80,35)(80,95){1}
\DashLine(80,95)(140,35){1}
\Vertex(120,35){2}\Text(120,25)[]{$x$}
%\Vertex(80,35){2}
\Vertex(80,95){2}\Text(80,105)[]{$\tilde x$}
\Vertex(140,35){2}
\end{picture}}
\qquad
\subfigure[]{\label{fig:AdSSESU}%
\begin{picture}(200,180)(0,0)\scriptsize
\SetWidth{1}
%\Boxc(100,90)(200,180)
\SetOffset(-200,0)
\SetScale{1}
\SetWidth{0}
\GBox(240,20)(360,160){0.8}
\SetWidth{1}
%\Line(240,25)(360,25)
\Text(240,10)[]{$\pi$}
\Text(270,10)[]{$\frac{\pi}{2}$}
\Text(300,10)[]{$\bar\rho=0$}
\Text(330,10)[]{$\frac{\pi}{2}$}
\Text(360,10)[]{$\pi$}
\Text(270,175)[]{$\omega<0$}
\Text(255,165)[]{$Y<0$}
\Text(330,175)[]{$\omega>0$}
\Text(300,165)[]{$Y>0$}
\Text(345,165)[]{$Y<0$}
\Line(240,20)(240,160)
\Line(270,20)(270,160)
\Line(300,20)(300,160)
\Line(330,20)(330,160)
\Line(360,20)(360,160)
\DashLine(240,35)(260,35){4}\Text(230,35)[cr]{$t$}
\DashLine(340,35)(360,35){4}
\DashLine(240,95)(260,95){4}\Text(230,95)[cr]{$t+\pi$}
\DashLine(280,95)(360,95){4}
\DashLine(260,95)(280,95){1}
\DashLine(320,35)(260,35){1}
\DashLine(260,35)(260,95){1}
\DashLine(260,95)(320,35){1}
\DashLine(320,35)(340,35){1}
\DashLine(280,35)(280,95){1}
\Vertex(320,35){2}\Text(320,25)[]{$z$}
\Vertex(260,95){2}\Text(260,105)[]{$\tilde z$}
\DashLine(280,95)(340,35){1}
\Vertex(280,95){2}\Text(280,105)[]{$\tilde z_\text{s}$}
\Vertex(340,35){2}\Text(340,25)[]{$z_\text{s}$}
\end{picture}}
\caption{$\AdS_2$ (Fig.\ \ref{fig:AdSESU}) and
  $\AdS_2\times\text{S}^0$ (Fig.\ \ref{fig:AdSSESU}) conformally mapped to the
  corresponding ESU. The regions that are covered are displayed as gray-filled
  regions. The ESU is given by a cylinder such
that one has to identify the two boundaries of the strip where $\rho=\pi$.
The two points of the $S^0$ within $\AdS_2$ and of the extra factor $S^0$ in
  the product space are $\omega=\pm1$ and $\frac{Y}{R}=\pm1$, respectively.
$\tilde x$ and $\tilde z$, $z_\text{s}$, $\tilde z_\text{s}$ are 
the antipodal points to $x$ and $z$ in respectively $\AdS_2$ and
$\AdS_2\times\text{S}^0$. They are 
constructed by following the lines with small dashsize. The horizontal 
direction
corresponds to the transformation in the space coordinates and the vertical 
one is associated to the time shift. The diagonal lines then point to the 
source at the corresponding conjugate point where null geodesics intersect. 
The conjugate points can be regarded as effective time shifted sources with
  the same time coordinate as the original event $x$ or $z$.}
\label{fig:antipodalpntinESU}
\end{center}
\end{figure}
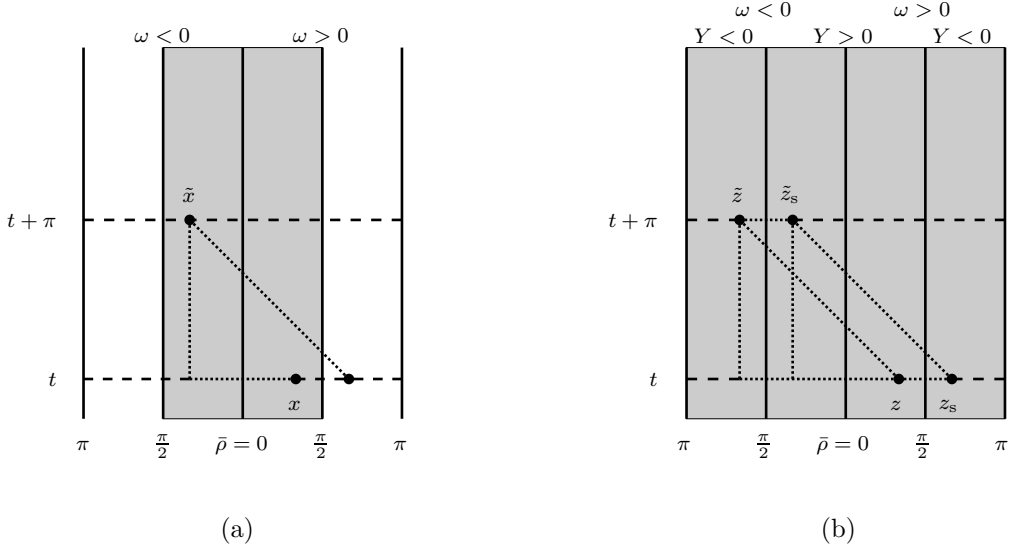

We now want to visualize the above relation on the sphere of the ESU. 
For convenience we choose $\AdS_2$ such that the ESU has topology
$\mathds{R}\times\text{S}^1$. The subsphere of $\AdS_2$ is given by
$\text{S}^0=\{-1,1\}$ such that we have $\omega=\pm1$. 
Hence, the transformation of $x_\Omega$ as prescribed in
\eqref{AdSantipodglobalcoord} becomes a flip between the two points of the
$\text{S}^0$. The information contained in $\text{S}^0$ can be traded for an 
additional sign information of $\bar\rho$, and therefore 
the transformation from  $x_\Omega$ to $\tilde x_\Omega$ 
simply corresponds to an reflection at $\bar\rho=0$. 
We will now describe the time shift. 
After the transformation of the spatial coordinates
is performed, one has found the antipodal event at time $t+\pi$. 
To relate it to an event at the original time $t$ one simply travels back
in time along any null geodesics that crosses the spatial position of the
antipodal event. On the ESU these null geodesics are clearly great circles.
They meet at two points on the sphere. One is at the spatial position of the 
event and the other point is the antipodal point on the sphere of the ESU.
The time it takes for a massless particle to travel between these two points 
is given by $\pi$, see Fig.\ \ref{fig:antipodalpntinESU}. 
In this way one now arrives at an event that can have caused the event at later
time $t+\pi$, and that has the same time coordinate as $x$, and its 
coordinate value $\bar\rho$ is given by a reflection at 
$\bar\rho=\frac{\pi}{2}$ on $\text{S}^1$. 
As $\bar\rho=\frac{\pi}{2}$ is the position of the AdS boundary, the mirror 
image to $x$ is situated outside of the region that corresponds to AdS. 
The effect of the original source at $x$ in combination with 
the mirror source either at $\tilde x$ as given in
\eqref{AdSantipodglobalcoord} or 
at equal times mirrored at the boundary is that a light ray that travels to the
boundary of AdS is reflected back into the interior.

Let us now discuss what happens in the case of $\AdSS$. The point
$z=(t,\bar\rho,x_\Omega,y)$ possesses 
the total antipodal point $\tilde z$ and the two semi-antipodal points 
$z_\text{s}$ and $\tilde z_\text{s}$ given by
\begin{equation}\label{AdSSantipodglobalcoord}
%\begin{aligned}
\tilde z=(t+\pi,\bar\rho,\tilde x_\Omega,\tilde y)\col\qquad
z_\text{s}=(t,\bar\rho,x_\Omega,\tilde y)\col\qquad
\tilde z_\text{s}=(t+\pi,\bar\rho,\tilde x_\Omega,y)\col
%\end{aligned}
\end{equation}
where $x_\Omega$ is as in the pure $\AdS_{d+1}$ case and fulfills 
\eqref{AdSsubsphereantipodalrel} and $y$ are all angle coordinates of 
$S^{d'+1}$.
%For the embedding coordinates $\vec Y$ of $S^{d'+1}$ one finds the relation
%\begin{equation}\label{Sphereantipodalrel}
%\vec Y(\tilde y)=-\vec Y(y)\pnt
%\end{equation}

In Fig.\ \ref{fig:antipodalpntinESU} the case of $\AdS_2\times\text{S}^0$, 
is shown. The effect of the factor $\text{S}^0$ can be alternatively described
by adding to the range $0\le\bar\rho\le\frac{\pi}{2}$ the interval  
$\frac{\pi}{2}\le\bar\rho\le\pi$. This is possible because in the ESU at
$\bar\rho=\frac{\pi}{2}$ the $\text{S}^0$ shrinks to a point.
The complete ESU is now covered by the image of 
$\AdS_2\times\text{S}^0$. 
The map to an antipodal position within the $\AdS_2$ factor is as before, 
one finds the spatial coordinates by reflecting at $\bar\rho=0$. 
Within the $\text{S}^0$ factor, the antipodal position is found by reflecting
at $\bar\rho=\frac{\pi}{2}$.
Using this, it can be seen that w.r.t.\ the point $z$,
the point $\tilde z$ is at the antipodal position on the $\text{S}^1$ of 
the ESU.
Traveling back in time from $t+\pi$ to $t$ along a null geodesic,
one arrives at $z$ from where one started. 
In the same way, the two semi-antipodal points 
$z_\text{s}$, $\tilde z_\text{s}$ are connected with each other by light
rays. On the sphere of the ESU the $z$ and $z_\text{s}$ are
related by a reflection at $\bar\rho=\frac{\pi}{2}$. Here, in contrast to 
the case of $\AdS_2$, even the mirror events at equal times are 
situated within the image of $\AdS_2\times\text{S}^0$. The above results 
are straightforwardly generalized to arbitrary dimensions.

Coming back to the discussion in Section \ref{weylana}, we can now make more 
precise statements about the mirror charge method to impose definite
boundary conditions at $\bar\rho=\frac{\pi}{2}$. A linear combination of 
the two solutions like in \eqref{AdSsuperpos} does not necessarily 
generate additional   
$\delta$-sources on the r.h.s.\ of the differential equation
\eqref{waveeq}, although both powerlike solutions in
\eqref{AdSsuperpos} have singularities within $\AdS_{d+1}$, the expression in
the first line has one at $x=x'$ and the expression in the second line has one
at $x=\tilde x'$. The singularity of the second expression only appears at
$t=t'+\pi$, and its contribution to the r.h.s.\ of the
differential equation \eqref{waveeq}
depends on the time ordering prescription.
In the cases where the $\theta$-function used for time ordering has
an additional step at $t=t'+\pi$, a second  $\delta$-function is generated
(see \cite{Avis:1978yn} for a discussion of $\AdS_4$). 
With the standard time ordering one finds that
$G_{\Delta_\pm}$ are solutions with a source at $x=x'$ only.
For $\AdS_4$ this was obtained in \cite{Dullemond:1985bc}.

The situation is different for $\AdSS$, where the propagator
\eqref{AdSSprop} has singularities at $z=z'$, $z=\tilde z'$ and 
the second solution \eqref{AdSSmirprop} has singularities at
$z=z'_\text{s}$, $z=\tilde z'_\text{s}$. 
Again, whether 
the singularities at $z=\tilde z'$ and $z=\tilde z'_\text{s}$ appear 
as $\delta$-sources 
on the r.h.s.\ of the differential equation \eqref{waveeq}, depends 
on the chosen time ordering. However in contrast to the pure $\AdS_{d+1}$
case, the singularity of the second solution \eqref{AdSSmirprop} at
$z=z'_\text{s}$ always leads to a $\delta$-source on the 
r.h.s.\ of \eqref{waveeq} but at the wrong position.
This result corresponds to the above observation on the ESU that the mirror
sources at equal times are not part of the image of $\AdS_{d+1}$ but of 
$\AdSS$. 

At the end let us give some comments on the
$i\varepsilon(t,t')$-prescription. First of all, one has to introduce it in all
expressions \eqref{AdSSprop}, \eqref{AdSSmirprop} and \eqref{AdSsuperpos},
since all of them have singularities at coincident or antipodal positions.
Secondly, as worked out for $\AdS_4$,
a time independent $\varepsilon(t,t')=\epsilon$ refers to taking the step
function $\theta(\sin(t-t'))$ for time ordering \cite{Avis:1978yn} which is 
appropriate if one restricts oneself to the hyperboloid. 
Standard time ordering with $\theta(t-t')$, being appropriate on the universal
covering, yields a time dependent
$\varepsilon(t,t')=\epsilon\sgn((t-t')\sin(t-t'))$ \cite{Dullemond:1985bc}.
As mentioned in Section \ref{waveeqapp}, due to the time dependence of
$\varepsilon(t,t')$, the coordinate dependence of the solutions is not entirely
included in $u$ and $v$.

%%%%%%%%%%%%%%%%%%%%%%%%%%%%%%%%%%%%%%%%%%%%%%%%%
%%%%%%%%%%%%%%%%%%%%%%%%%%%%%%%%%%%%%%%%%%%%%%%%%%
\section{Mode summation on $\AdSS$}
\label{modesummation}
In this Section we will use the propagator on
pure $\text{AdS}_{d+1}$ given by \eqref{AdSprop}
and the spherical harmonics on $\text{S}^{d'+1}$ to construct the
propagator on $\AdSS$ via its mode expansion, summing up all the KK modes. 
We will be able to perform the sum only
for special mass values where the conformal dimensions $\Delta_\pm$ of the
scalar modes are linear functions of $l$, with $l$ denoting the $l$th  mode in
the KK tower. Even a mixing of several scalar modes of this kind is allowed. 
The mixing case is interesting because it occurs in
supergravity theories on $\AdS_{d+1}\times\text{S}^{d+1}$ backgrounds
\cite{Corley:1999uz,Michelson:1999kn,Deger:1998nm,Kim:1985ez,Gunaydin:1985fk}.
For example in type $\twob$ supergravity in $\text{AdS}_5\times\text{S}^5$ the
mass eigenstates of the mixing matrix for scalar modes
\cite{Kim:1985ez,Gunaydin:1985fk} correspond to the bosonic
chiral primary and descendant operators in the $\AdS/\text{CFT}$ dictionary
\cite{Lee:1998bx}.
For these modes  $\Delta_\pm$ depend linearly on $l$.

The main motivation for investigating the mode summation was the hope to find
the propagator for generic mass values. But forced to stay in a regime of a
linear $\Delta_\pm$ versus $l$ relation we can give up the condition of
conformal flatness, but remain restricted to special mass values. We
nevertheless present this study since several interesting aspects are found
along the way. Furthermore, in the literature it is believed that an explicit
computation of the KK mode summation is too cumbersome \cite{Mathur:2002ry}. 
We will show how to deal with the mode summation by discussing the
$\AdS_3\times\text{S}^3$ case first, allowing for unequal radii but
necessarily a special mass value. The result will then be compared to the
expressions in the previous Sections by specializing to equal embedding
radii. 

Having discussed this special case we will comment on the modifications which
are necessary to deal with generic $\AdSS$ spacetimes.  

The results of the previous Sections in connection with the expression for the
mode summation in the conformally flat and Weyl invariant coupled case lead to
the formulation of a summation rule for a product of Legendre functions and
Gegenbauer polynomials. An independent proof of this rule is given in Appendix
\ref{userel}. With this it is possible to discuss the 
results in generic dimensions without doing all the computations
explicitly. Furthermore, the sum rule might be useful for other applications,
too.  

For the solution of \eqref{waveeq} we make the following 
ansatz\footnote{This ansatz is designed to generate a solution that 
corresponds to \eqref{AdSSprop}. If one wants to generate a solution
corresponding to \eqref{AdSSmirprop} one has to replace either $y$ or $y'$ by
the corresponding antipodal coordinates $\tilde y$ or $\tilde y'$.} 
\begin{equation}\label{AdSSpropmodeexp}
G(z,z')=\frac{1}{R_2^{d'+1}}\sum_{I}G_I(x,x')Y^I(y)Y^{\ast I}(y')\col
\end{equation} 
where we sum over the multiindex $I=(l,m_1,\dots,m_{d'})$ such that $l\ge
m_1\ge\dots\ge m_{d'-1} \ge | m_{d'}|\ge 0$, $Y^I$ denote the spherical
harmonics on $\text{S}^{d'+1}$, and `$\ast$' means complex conjugation. 
Some useful relations for the spherical
harmonics can be found in Appendix \ref{userel}.

The mode dependent Green function on $\AdS_{d+1}$ then fulfills
\begin{equation}
\Big(\Box_x
-M^2-\frac{l(l+d')}{R_2^2}\Big)G_I(x,x')=\frac{i}{\sqrt{-g_\text{AdS}}}\delta
(x,x')\col 
\end{equation} 
which follows when decomposing the d'Alembert operator like in
\eqref{dalemchordal} and using \eqref{SHcasimir}. 
The solution of this equation was already given in \eqref{AdSprop}, into which
the (now KK mode dependent) conformal dimensions enter. They were already
defined in \eqref{conformaldim}, and the AdS mass is a function of the mode
label $l$ 
\begin{equation}\label{mkksphere}
m^2=M^2+m_\text{KK}^2=M^2+\frac{l(l+d')}{R_2^2}\pnt
\end{equation}

In the following as a simple example we will present the derivation of the propagator on $\AdS_3\times\text{S}^3$ via the KK mode summation. Compared to the physically more interesting $\AdS_5\times\text{S}^5$ background the expressions are easier and the general formalism becomes clear.

Evaluating \eqref{AdSprop} for $d=d'=2$ the $\AdS_3$ propagator for the $l$th
KK mode is given by 
\begin{equation}\label{AdS3propexpl}
G_\Delta(x,x')=\frac{1}{R_12^{\Delta+1}\pi}\xi^\Delta\hypergeometric{\tfrac{\Delta}{2}}{\tfrac{\Delta}{2}+\tfrac{1}{2}}{\Delta}{\xi^2}=\frac{1}{R_14\pi}\frac{1+\sqrt{1-\xi^2}}{\sqrt{1-\xi^2}}\Big[\frac{\xi}{1+\sqrt{1-\xi^2}}\Big]^\Delta\pnt
\end{equation}
From \eqref{hypergeomstart}, \eqref{conformaldim} and \eqref{mkksphere} one
finds that the mode dependent positive branch of the conformal dimension reads 
\begin{equation}\label{AdS3S3conformaldim}
\Delta=\Delta_+=1+\frac{R_1}{R_2}\sqrt{\frac{R_2^2}{R_1^2}+l(l+2)+M^2R_2^2}\pnt
\end{equation}

The spherical part follows from \eqref{SHcomplete} of Appendix
\ref{userel} where we discuss it in more detail and is given by
\begin{equation}
\sum_{\scriptscriptstyle m_1\ge| m_2|\ge0 } ^l Y^I(y) Y^{\ast I}(y') = 
\frac{(l+1)}{2 \pi^2}
C_l^{(1)}(\cos\Theta)\col\qquad\cos\Theta=\frac{Y\cdot
  Y'}{R_2^2}=1-\frac{v}{2R_2^2}\pnt  
\end{equation}
Remember that the $C_l^{(\beta)}$ denote the Gegenbauer polynomials and $Y$,
$Y'$ in the formula for $\Theta$ are the embedding space coordinates of the
sphere, compare with \eqref{defeq} and \eqref{uv}. 
One thus obtains from \eqref{AdSSpropmodeexp}
\begin{equation}
G(z,z')=\frac{1}{8\pi^3R_1R_2^3}\frac{1+\sqrt{1-\xi^2}}{\sqrt{1-\xi^2}}\sum_{l=0}^\infty(l+1)\Big[\frac{\xi}{1+\sqrt{1-\xi^2}}\Big]^\Delta
C_l^{(1)}(\cos\Theta)\pnt 
\end{equation} 
In this formula $\Delta$ is a function of the mode parameter $l$ and we can
explicitly perform the sum only for special conformal dimensions which are
linear functions of $l$ 
\begin{equation}
\Delta=\Delta_+=\frac{R_1}{R_2}l+\frac{R_1+R_2}{R_2}\col
\end{equation}
following from \eqref{AdS3S3conformaldim} after choosing the special mass value
\begin{equation}\label{AdS3S3specmass}
M^2=\frac{1}{R_2^2}-\frac{1}{R_1^2}\pnt
\end{equation} 

The sum then simplifies and can explicitly be evaluated by a reformulation of
the $l$-dependent prefactor as a derivative and by using \eqref{genfunc} 
\begin{equation}\label{sumcalc}
\sum_{l=0}^\infty(l+1)q^l
C_l^{(1)}(\eta)=\Big(q\preparderiv{q}+1\Big)\sum_{l=0}^\infty q^l
C_l^{(1)}(\eta)=\frac{1-q^2}{(1-2q\eta+q^2)^2}\pnt 
\end{equation}
With the replacements
\begin{equation}\label{qetadef}
q=\Big[\frac{\xi}{1+\sqrt{1-\xi^2}}\Big]^\frac{R_1}{R_2}\col\qquad
\eta=\cos\Theta 
\end{equation}
one now finds after some simplifications
\begin{equation}
G(z,z')=\frac{1}{8\pi^3R_1R_2^3}\frac{1}{\sqrt{1-\xi^2}}\xi^{1+\frac{R_1}{R_2}}\frac{(1+\sqrt{1-\xi^2})^\frac{R_1}{R_2}-(1-\sqrt{1-\xi^2})^\frac{R_1}{R_2}}{\Big[(1+\sqrt{1-\xi^2})^\frac{R_1}{R_2}-2\xi^\frac{R_1}{R_2}\cos\Theta+(1-\sqrt{1-\xi^2})^\frac{R_1}{R_2}\Big]^2}\pnt
\end{equation}
For the conformally flat case $R_1=R_2=R$, where \eqref{AdS3S3specmass}
becomes the mass generated by the Weyl invariant coupling to the background,
the above expression simplifies to 
\begin{equation}
G(z,z')=\frac{1}{4\pi^3R^4}\frac{\xi^2}{(2-2\xi\cos\Theta)^2}=\frac{1}{4\pi^3}\frac{1}{(u+v+i\varepsilon(t,t'))^2}\col
\end{equation}
where we have restored the $i\varepsilon(t,t')$-prescription. 
This result exactly matches \eqref{AdSSprop}.

The way to perform the KK mode summation on generic $\AdSS$ backgrounds is
very similar to the one presented above. 
One finds a linear relation between $l$ and $\Delta$ 
\begin{equation}
\Delta_\pm=\pm\frac{R_1}{R_2}l+\frac{dR_2\pm d'R_1}{2R_2}
\end{equation}
at the $(d+d'+2)$-dimensional mass value
\begin{equation}
M^2=\frac{d'^2R_1^2-d^2R_2^2}{4R_1^2R_2^2}\pnt
\end{equation}
This expression is a generalization of \eqref{AdS3S3specmass} and it 
reduces to \eqref{conditions} in the conformally flat case.
For generic dimension the way of computing the propagator is very similar to
the one presented for the $\AdS_3\times\text{S}^3$ background. However the
steps \eqref{AdS3propexpl} to express the hypergeometric function in the AdS
propagator and \eqref{sumcalc} to compute the sum become more tedious. For
dealing with the hypergeometric functions see the remarks in Appendix
\ref{userel}. 
The sum generalizes in the way, that higher derivatives and more terms enter 
the expression \eqref{sumcalc}. \\ 

Next we discuss the mode summation in the conformally flat case $R_1=R_2$ at
the Weyl invariant mass value but for generic $d$ and $d'$. 
In this case with the corresponding conformal dimensions
\begin{equation}\label{conformaldimconf}
\Delta=\Delta_+=l+\frac{d+d'}{2}\col
\end{equation}
using \eqref{AdSprop} and \eqref{SHcomplete}, the propagator is expressed as
\begin{equation}\label{modesumprop}
\begin{aligned}
G(z,z')&=\frac{\Gamma(\frac{d'}{2})}{4\pi}\Big(\frac{\xi}{2\pi
  R^2}\Big)^{\frac{d+d'}{2}}\\
&\phantom{={}}
\times\sum_{l=0}^\infty\frac{\Gamma(l+\frac{d+d'}{2})}{\Gamma(l+\frac{d'}{2})}\Big(\frac{\xi}{2}\Big)^l\hypergeometric{\tfrac{l}{2}+\tfrac{d+d'}{4}}{\tfrac{l}{2}+\tfrac{d+d'}{4}+\tfrac{1}{2}}{l+\tfrac{d'}{2}+1}{\xi^2}C_l^{(\frac{d'}{2})}(1-\tfrac{v}{2R^2})\pnt
\end{aligned}
\end{equation} 
This equality together with the solution \eqref{AdSSprop} has lead us to
formulate a sum rule for the above given functions at generic $d$ and
$d'$. The above series should exactly reproduce \eqref{AdSSprop}. 
In Appendix \ref{userel} we give an independent direct proof of the sum rule. 

Considering the mode summation one finds an interpretation of the asymptotic
behaviour of \eqref{AdSSprop} observed in Subsection \ref{AdSSderiv}. 
The asymptotic regime $u\to\infty$ corresponds to $\xi\to0$. As the
contribution of the $l$th mode is proportional to $\xi^{\Delta_+}\sim\xi^l$,
the conformal dimension of the zero mode determines the asymptotic behaviour.  

Note also that the additional singularity of \eqref{AdSSprop} at the total
antipodal position $z=\tilde z'$ can be seen already in
\eqref{AdSSpropmodeexp}. 
Under antipodal reflection in $\AdS_{d+1}$ the pure AdS propagator fulfills
$G_{\Delta_\pm}(x,\tilde x')=(-1)^{\Delta_\pm}G_{\Delta_\pm}(x,x')$. On the 
sphere the spherical harmonics at antipodal points are related via
$Y^I(y)=(-1)^lY^I(\tilde y)$. Hence, in case that $\Delta_\pm$ is given by
\eqref{conformaldimconf}, replacing $z'$ by the total antipodal point 
$\tilde z'$ leads to the same expression for the mode sum up to an
$l$-independent phase factor. 

One final remark to the choice of $\Delta_+$. What happens if one performs the
mode expansion with AdS propagators based on $\Delta_-$? First in any case for
high enough KK modes $\Delta_-$ violates the unitarity bound
\eqref{splitbound}. But ignoring this condition from physics one can
nevertheless study the mathematical issue of summing with $\Delta_-$. The
corresponding series is given by \eqref{sumcalc} after replacing $q$ by
$q^{-1}$. It is divergent since for real $u$ the variable $q$ in
\eqref{qetadef} obeys $|q|\le 1$ (case $R_1=R_2$). One can give meaning to the
sum by the following procedure. $q$ as a function of $\xi$ has a cut between
$\xi=\pm 1$. If $|q|\le 1$ on the upper side of the cut, then $|q|\ge 1$ on
the lower side. Hence it is natural to define the sum with $\Delta_-$ as the
analytic continuation from the lower side. By this procedure we found both for
$\AdS_3\times\text{S}^3$ and  $\AdS_5\times\text{S}^5$ up to an overall factor
$-1$ the same result as using $\Delta_+$. The sign factor can be understood as
a consequence of the continuation procedure.  

%%%%%%%%%%%%%%%%%%%%%%%%%%%%%%%%%%%
%%%%%%%%%%%%%%%%%%%%%%%%%%%%%%%%%%%%%%
\section{The plane wave limit}
\label{pwlimit}
The  plane wave background arises as a certain Penrose limit of
$\AdS_5\times\text{S}^5$. The scalar propagator in the plane wave has been
constructed in \cite{Mathur:2002ry}. In this Section we study how this
propagator in the massless case arises as a limit of our
$\AdS_5\times\text{S}^5$ propagator \eqref{AdSSprop}.   

This approach is in the spirit of \cite{Dorn:2003ct}, where one follows the
limiting process instead of taking the limit before starting any
computations. One finds a simple interpretation of certain functions of the
coordinates introduced in \cite{Mathur:2002ry}. 

As an additional consistency check we take the $R\to\infty$ limit of the
differential equation \eqref{waveeq} using \eqref{dalemchordal} to obtain the
equation on the plane wave background and find that it is fulfilled by the
massive propagator given in \cite{Mathur:2002ry}. 

Taking the aforementioned Penrose limit of $\AdS_5\times\text{S}^5$ means to
focus into the neighbourhood of a certain null geodesic which runs along an
equator of the sphere with velocity of light. The metric of
$\AdS_5\times\text{S}^5$ in global coordinates\footnote{The $\AdS_5$
  coordinates are related to the ones in \eqref{AdSmetricglobalcoord} via
  $\cosh\rho=\sec\bar\rho$.}  
\begin{equation}
\de s^2=R^2\big(-\de t^2\cosh^2\rho+\de\rho^2+\sinh^2\rho
\de\Omega_3^2+\de\psi^2\cos^2\vartheta+\de\vartheta^2+\sin^2\vartheta
\de\hat\Omega_3^2\big) 
\end{equation}
via the replacements
\begin{equation}\label{vartraf}
t=z^++\frac{z^-}{R^2}\col\qquad\psi=z^+-\frac{z^-}{R^2}\col\qquad\rho=\frac{r}{R}\col\qquad\vartheta
=\frac{y}{R} 
\end{equation}
in the $R\to\infty$ limit turns into the plane wave metric.

The relation between global coordinates and the embedding space coordinates,
see e.~g.~\cite{Aharony:1999ti}, yields ($\omega_i$, $\hat\omega_i$,
$i=1,\dots,4$ with $\vec\omega^2=\vec{\hat\omega}^2=1$ are the embedding
coordinates of the two unit 3-spheres) 
\begin{equation}
\begin{aligned}
u&=2R^2\Big[-1+\cosh\rho\cosh\rho'\cos(t-t')-\sinh\rho\sinh\rho'\,\omega_i\omega'_i\Big]\\  
v&=2R^2\Big[+1-\cos\vartheta\cos\vartheta'\cos(\psi-\psi')-\sin\vartheta\sin\vartheta'\,\hat\omega_i\hat\omega'_i\Big]\pnt 
\end{aligned}
\end{equation} 
Applying \eqref{vartraf} one gets at large $R$ up to terms vanishing for
$R\to\infty$ 
\begin{equation}\label{chordalpw}
\begin{aligned}
u&=2R^2\Big[-1+\cos\Delta z^++\frac{1}{R^2}\Big(-(\vec x^2+\vec
x{\hspace{0.5pt}}'^2)\sin^2\frac{\Delta z^+}{2}+\frac{(\vec x-\vec
  x{\hspace{0.5pt}}')^2}{2}-\Delta z^-\sin\Delta z^+\Big)\Big]\\ 
v&=2R^2\Big[+1-\cos\Delta z^++\frac{1}{R^2}\Big(-(\vec y^2+\vec
y{\hspace{1pt}}'^2)\sin^2\frac{\Delta z^+}{2}+\frac{(\vec y-\vec
  y{\hspace{1pt}}')^2}{2}-\Delta z^-\sin\Delta z^+\Big)\Big]\col 
\end{aligned}
\end{equation}
where $\Delta z^\pm=z^\pm-z'^\pm$ and $\vec x=r\vec\omega$, $\vec
y=y\vec{\hat\omega}$. 
In the $R\to\infty$ limit the sum of both chordal distances is thus given by
\begin{equation}\label{totalchordalpw}
\Phi=\lim_{R\to\infty}(u+v)=-2(\vec z^2+\vec
z{\hspace{1pt}}'^2)\sin^2\frac{\Delta z^+}{2}+(\vec z-\vec
z{\hspace{1pt}}')^2-4\Delta z^-\sin\Delta z^+\col
\end{equation}
where $\vec z=(\vec x,\vec y)$, 
$\vec z{\hspace{1pt}}'=(\vec x{\hspace{0.5pt}}',\vec y{\hspace{1pt}}')$ and
$\Phi$ refers 
to the notation of \cite{Mathur:2002ry}. $\Phi$ is precisely the $R\to\infty$
limit of the total chordal distance on $\AdS_5\times\text{S}^5$, which remains
finite as both $\sim R^2$ terms in \eqref{chordalpw} cancel. This happens due
to the expansion around a \emph{null} geodesic.
In Appendix \ref{chordaldistpw} it is shown that $\Phi$ is the chordal
distance in the plane wave.

The massless propagator in the plane wave background in the $R\to\infty$ limit
of \eqref{AdSSprop} with $d=d'=4$ thus becomes 
\begin{equation}
G_\text{pw}(z,z')=\frac{3}{2\pi^5}\frac{1}{(\Phi+i\varepsilon(z^+,z{\hspace{1pt}}'^+))^4}\col
\end{equation}  
which agrees with \cite{Mathur:2002ry}. 

In addition we checked the massive propagator of \cite{Mathur:2002ry} which
fulfills the differential equation on the plane wave background. This equation
can be obtained from \eqref{waveeq} and \eqref{dalemchordal} by taking the
$R\to\infty$ limit. In the limit the sum of both chordal distances is given in
\eqref{totalchordalpw}. The difference is given by 
\begin{equation}\label{diffchordalpw}
\lim_{R\to\infty}\frac{u-v}{R^2}=4(\cos\Delta z^+-1)
\end{equation}
this has to be substituted into \eqref{dalemchordal}. 
Finally, one obtains the differential equation
\begin{equation}
\Big[4\cos\Delta
z^+\Big(5\preparderiv{\Phi}+\Phi\doublepreparderiv{\Phi}\Big)+4\sin\Delta
z^+\preparderiv{\Phi}\preparderiv{\Delta
  z^+}-M^2\Big]G_\text{pw}(z,z')=\frac{i}{\sqrt{-g_\text{pw}}}\delta
(z,z')\col
\end{equation}
which is fulfilled by the expression given in \cite{Mathur:2002ry}.
As already noticed in Section \ref{waveeqapp}, in contrast to the massless
propagator the massive one depends not only on the total chordal distance
$\Phi$ but in addition on \eqref{diffchordalpw}. 
%%%%%%%%%%%%%%%%%%%%%%%%%%%%%%%%%%%%
%%%%%%%%%%%%%%%%%%%%%%%%%%%%%%%%%%%%
\section{Conclusions}
\label{concl}
In this paper we have focussed on the propagator of scalar
fields on $\AdSS$ backgrounds. 
We have
argued that for an investigation of holography in the plane wave, 
in a first step one should study
this propagator instead of the bulk-to-boundary 
one, since only for the former the Penrose limit is well defined. 

We have first discussed solutions of the defining propagator
equation at points away from possible singularities.
On conformally flat backgrounds and for Weyl invariant coupled fields,
both in $\AdSS$ and in pure $\AdS_{d+1}$, exist two solutions, which are
powerlike in the chordal distances. For $\AdSS$ these two solutions are powers
either in the sum or difference of the chordal distances with respect to the
$\AdS_{d+1}$ and $\text{S}^{d'+1}$ factor. The first solution has a
singularity 
if both points coincide or if they are at antipodal positions to each other. 
The second solution has singularities
where both points are semi-antipodal to each other. 

Whether, acting with the d'Alembert operator, $\delta$-sources are generated
at the locations of these singularities, depends on the time ordering
prescription. For $\AdS_{d+1}$, being the
universal cover of the embedded hyperboloid, standard time ordering is
appropriate. Then for both solutions source terms
arise only at coinciding times. This implies that the first solution
develops just the right source to solve the full propagator equation.
But the second solution necessarily has a source term away from the
coincidence of the two points. Therefore it cannot be used to form different
propagators via linear combinations with the first solution.

This is in contrast to the pure $\AdS_{d+1}$ case. There the second solution
has a singularity at the position where 
both points are antipodal to each other on the hyperboloid, i.e. there is no
singularity at coinciding time coordinates. 
Hence, this singularity does not lead to a $\delta$-source. 
Thus, taking the sum and difference of the two solutions, 
propagators, which obey respectively Neumann and Dirichlet boundary
conditions, can be constructed. 

In addition, for $\AdSS$ we have investigated the KK decomposition of the
propagator using spherical harmonics. We have noted that the summation can be
performed even in non conformally flat backgrounds, but only for special mass
values. The relevant condition is that the conformal dimension of the field
mode is a linear function of the KK mode parameter.  

In the conformally flat case for a Weyl invariant coupled field the uniqueness
of the solution of the differential equation in combination with the KK
decomposition led to the formulation of a theorem that sums up a product of
Legendre functions and Gegenbauer polynomials. It was independently proven 
in Appendix \ref{userel}. 

For $\AdS_5\times\text{S}^5$ we explicitly performed the Penrose
limit on our expression for the propagator to find the result on the plane
wave background. We found agreement with the literature
\cite{Mathur:2002ry} and got an interpretation for the spacetime
dependence of the result. It simply depends on the $R\to\infty$ limit of the
sum of both chordal distances on $\AdS_5\times\text{S}^5$, which 
was shown to be the chordal distance in the plane wave. In the
general massive case there is an additional dependence on the suitable
rescaled difference of both chordal distances.  
We formulated the differential equation in the limit and checked
that the massive propagator on the plane wave background given in
\cite{Mathur:2002ry} is a solution.

Clearly future work is necessary to construct the propagator for the case of
generic mass values. But already with our results one should be able to
address the issue of defining a bulk-to-boundary propagator
in the plane wave limit.\\ 

%%%%%%%%%%%%%%%%%%%%%%%%%%%%%%%%%%%%
%%%%%%%%%%%%%%%%%%%%%%%%%%%%%%%%%%%%
\noindent{\bf Acknowledgements}
\\The work was supported by DFG (German Science Foundation) with the
``Schwerpunktprogramm Stringtheorie" and the ``Graduiertenkolleg 271". 
We thank James Babington, Niklas Beisert, Danilo Diaz, Johanna Erdmenger, 
George Jorjadze and Nicolaos Prezas for useful discussions.  
%%%%%%%%%%%%%%%%%%%%%%%%%%%%%%%%%%%%
%%%%%%%%%%%%%%%%%%%%%%%%%%%%%%%%%%%%
\numberwithin{equation}{section}
\appendix
\section{Relation of the bulk-to-bulk and the bulk-to-bound\-ary propagator}
\label{bulkboundproprel}
We will show for a scalar field in an Euclidean space how the bulk-to-boundary
propagator is related to the bulk-to-bulk propagator, if the boundary has
codimension one w.r.t.\ the bulk, like in the case of the 
$\AdS/\text{CFT}$ correspondence.
The bulk-to-bulk propagator $G(x,x')$ of a scalar field with mass $m$ is
defined as Green function that fulfills\footnote{This equation is the analytic
continuation of \eqref{waveeq}.}
\begin{equation}\label{Gfuncdiffeq}
(\Box_x -m^2)G(x,x')=-\frac{1}{\sqrt{g}}\delta (x,x')\col
\end{equation} 
with appropriate boundary conditions. Here
$\Box_x$ is the Laplace operator on the $(d+1)$-dimensional Riemannian 
manifold $M$ with a $d$-dimensional boundary which we denote with 
$\partial M$. The
coordinates are $x^i$, the metric is $g_{ij}$ and its determinant is $g$. The
  propagator $G(x,x')$ corresponds to a scalar field $\phi(x)$ which should
  obey 
\begin{equation}\label{phidiffeq}
(\Box_x -m^2)\phi(x)=J(x)\col\qquad\lim_{x_\perp\to 0}\phi(x)x^a_\perp=\bar\phi(\bar x)\col
\end{equation}
where $J(x)$ are sources for the field $\phi$ in the interior. We have split
the coordinates like $x=(x_\perp,\bar x)$ with the boundary at
$x_\perp=0$. Boundary values $\bar\phi$ for the field $\phi$ 
are specified with a
nontrivial scaling with $x^a_\perp$ for later convenience. 
The bulk-to-boundary propagator $K(x,\bar x')$ is defined as the solution of 
the equations
\begin{equation}\label{bulkboundpropdefeq}
(\Box_x -m^2)K(x,\bar x')=0\col\qquad\lim_{x_\perp\to 0}K(x,\bar
x')x^a_\perp=\delta(\bar x,\bar x')\col
\end{equation}
where the second equation implements the necessary singular behaviour at the
boundary. 
A solution of the equations \eqref{phidiffeq} 
with $J(x)=0$ is then given by
\begin{equation}\label{boundvalsol}
\phi(x)=\int_{\partial M}\de^d\bar x' K(x,\bar x')\bar\phi(\bar x')\pnt
\end{equation}
Since we deal with the problem in Euclidean signature
\cite{Witten:1998qj,Freedman:1998tz,D'Hoker:2002aw},
we will denote $K(x,\bar x')$ as the Poisson kernel. It
is not independent from the Green function defined via \eqref{Gfuncdiffeq} as
we will now show. 

With \eqref{Gfuncdiffeq} one can write an identity for the field $\phi$ that
reads 
\begin{equation}
\phi(x)=-\int_M\de^{d+1}x'\sqrt{g}\phi(x')(\Box_{x'}-m^2)G(x,x')\pnt
\end{equation}
After applying partial integration twice and using \eqref{phidiffeq} it
assumes the form
\begin{equation}\label{boundsourcesol}
\phi(x)=\int_{\partial M}\de
A'_\mu\sqrt{g}g^{\mu\nu}\big[(\partial'_\nu\phi(x'))G(x,x')-\phi(x')\partial'_\nu
G(x,x')\big]-\int_M\de^{d+1}x'\sqrt{g}J(x')G(x,x')\col
\end{equation}
where $\de A'_\mu$ denotes the infinitesimal area element on $\partial M$ which
points into the outer normal direction and $\partial'_\mu$ denotes a derivative
w.r.t.\ $x'_\mu$. If one has the additional restriction that $G(x,x')=0$
for $x'\in\partial M$ ($x'_\perp=0$) the first term in the above boundary
integral is zero.  One then arrives at the `magic rule'  
which for the boundary value problem in presence of a source $J(x)$ in the
interior can be found in \cite{Barton:1989}.  
Here, however, we have to be more
careful. In \eqref{phidiffeq} we have allowed for a scaling of the boundary
value with $x^a_\perp$ as written down in \eqref{bulkboundpropdefeq}. For
$a>0$ the vanishing $G(x,x')$ at the boundary can be compensated and the 
first term in the boundary integral of \eqref{boundsourcesol} then
contributes. 

Considering $\AdS_{d+1}$, this is indeed the case, because the
field $\phi$ with conformal dimension $\Delta$ represents the non-normalizable 
modes, which scale as given in \eqref{bulkboundpropdefeq} with $a=\Delta-d$.
Indicating the corresponding propagator with the suffix $\Delta$, 
one finds $G_\Delta(x,x')=0$
at $x'\in\partial M$ but the vanishing is compensated by the singular
behaviour of the non-normalizable modes in the limit $x'_\perp\to 0$.

We now formulate \eqref{boundsourcesol} on Euclidean $\AdS_{d+1}$ in Poincar\'e
coordinates %\eqref{AdSmetricPcoord} (with the minus sign in front of $\de
            %x_0^2$ converted to plus) 
with $J(x')=0$, where one has
\begin{equation}
\de A_\mu=-\de^d\bar
x\delta_\mu^\perp\col\qquad\sqrt{g}=\Big(\frac{R_1}{x_\perp}\Big)^{d+1}\col\qquad g^{\perp\perp}=\Big(\frac{x_\perp}{R_1}\Big)^2\pnt
\end{equation}
The minus sign in the area element stems from the fact that the
$x_\perp$-direction points into the interior of $\AdS_{d+1}$, 
but one has to take
the outer normal vector. Now \eqref{boundsourcesol} reads
\begin{equation}
\phi_\Delta(x)=-R_1^{d-1}\int\de^d\bar
x'x'^{1-d}_\perp\big[(\partial'_\perp\phi(x'))G_\Delta(x,x')-\phi(x')\partial'_\perp G_\Delta(x,x')\big]\pnt
\end{equation}
Using now \eqref{bulkboundpropdefeq} with $a=\Delta-d$, one finds that the
relation of the bulk-to-bulk and the bulk-to-boundary propagator is given by 
\begin{equation}\label{bulktoboundarypropagatorrel}
K_\Delta(x,\bar x')=-R_1^{d-1}
\big[(d-\Delta)x'^{-\Delta}_\perp-x'^{1-\Delta}_\perp\partial'_\perp
\big] G_\Delta(x,x')\big|_{x'_\perp=0}
\end{equation}
If we now insert the explicit expression \eqref{AdSprop} for $G_\Delta(x,x')$, 
we see what we already mentioned: in approaching the boundary  
($x'_\perp\to 0$), $G_\Delta(x,x')$ itself goes to zero 
like $x'^\Delta_\perp$ but this is compensated by the singular behaviour of
the prefactor in the first 
term of \eqref{bulktoboundarypropagatorrel}.
Hence, in contrast to the situation of the 'magic rule' \cite{Barton:1989}, 
it contributes to the bulk-to-boundary propagator. 
One then finds with 
\begin{equation}\label{xiinPcoord}
\xi=\frac{2x_\perp x'_\perp}{x_\perp^2+x'^2_\perp+(\bar x-\bar x')^2}\pnt
\end{equation} 
in Poincar\'e coordinates and with $\hypergeometric{a}{b}{c}{0}=1$ that
effectively  
\begin{equation}
K_\Delta(x,\bar x')=-R_1^{d-1}(d-2\Delta)x'^{-\Delta}_\perp G_\Delta(x,x')\big|_{x'_\perp=0}\col
\end{equation}
which is in perfect agreement with the explicit expressions
given in \cite{D'Hoker:2002aw}.
%\eqref{bulkboundAdSprop} and \eqref{AdSprop}. 

%%%%%%%%%%%%%%%%%%%%%%%%%%%%%%%%%
%%%%%%%%%%%%%%%%%%%%%%%%%%%%%%%%%
\section{Useful relations for hypergeometric functions and spherical
  harmonics}
\label{userel} 
Most of the relations we present here can be found in
\cite{Bateman:1953,Abramowitz:1972,Gradshteyn:1980} or are derived from there.
The hypergeometric functions in the propagators \eqref{AdSprop} with
$\Delta_\pm=\frac{d\pm1}{2}$ (at the  mass value generated by the Weyl
invariant coupling) become ordinary analytic expressions  
\begin{equation}\label{hypergeom1}
\begin{aligned}
\hypergeometric{a}{a+\tfrac{1}{2}}{\tfrac{1}{2}}{\xi^2}&=\frac{1}{2}\big[(1+\xi)^{-2a}+(1-\xi)^{2a}\big]\col\\ 
\hypergeometric{a+\tfrac{1}{2}}{a+1}{\tfrac{3}{2}}{\xi^2}&=-\frac{1}{4a\xi}\big[(1+\xi)^{-2a}-(1-\xi)^{-2a}\big]\pnt 
\end{aligned}
\end{equation}
Setting $a=\frac{d-1}{4}$ one finds \eqref{AdSsuperpos}.

To find the hypergeometric functions relevant for the propagators in higher
dimensional AdS spaces one can use a recurrence relation (Gau\ss' relation for
contiguous functions)   
\begin{equation}\label{recurrencerel}
\hypergeometric{a}{b}{c-1}{z}=\frac{c[c-1-(2c-a-b-1)z]}{c(c-1)(1-z)}\hypergeometric{a}{b}{c}{z}+\frac{(c-a)(c-b)z}{c(c-1)(1-z)}\hypergeometric{a}{b}{c+1}{z}\pnt
\end{equation}
where the hypergeometric functions relevant in lower dimensional AdS spaces
enter. 

For odd AdS dimensions (even $d$) the relevant hypergeometric functions can be
expressed with the above recurrence relation in terms of ordinary analytic
functions. This happens because of the explicit expressions 
\begin{equation}
\begin{aligned}\label{hypergeomstart}
\hypergeometric{a}{a+\tfrac{1}{2}}{2a}{z}&=\frac{2^{2a-1}}{\sqrt{1-z}}\big[1+\sqrt{1-z}\big]^{1-2a}\col\\ 
\hypergeometric{a}{a+\tfrac{1}{2}}{2a+1}{z}&=2^{2a}\big[1+\sqrt{1-z}\big]^{-2a}\pnt 
\end{aligned}
\end{equation}
One has to apply \eqref{recurrencerel} $n$ times to compute the AdS propagator
at generic $\Delta$ in $d+1=3+2n$ dimensions. For $\AdS_3$ one simply uses the
first expression in \eqref{hypergeomstart}. 
The $\AdS_5$ case is of particular importance and therefore we give the
explicit expression for the needed hypergeometric function 
\begin{equation}
\hypergeometric{\tfrac{\Delta}{2}}{\tfrac{\Delta}{2}+\tfrac{1}{2}}{\Delta-1}{z}=\frac{1}{2(1-z)^\frac{3}{2}}\Big[\frac{2}{1+\sqrt{1-z}}\Big]^{\Delta-1}\Big[\sqrt{1-z}+\frac{\Delta-1}{\Delta-2}(1-z)+\frac{1}{\Delta-1}\Big]\pnt 
\end{equation}

Spherical harmonics $Y^I(y)$ on $\text{S}^{d'+1}$ are characterized by quantum
numbers 
\begin{equation}\label{multindex}
I = (l, m_1, \dots, m_{d'})\col\qquad l  \ge m_1 \ge \dots \ge m_{d'-1} \ge |
m_{d'}| \ge 0 
\end{equation}
 and form irreducible representations of $\text{SO}(d'+2)$. They are
 eigenfunctions with respect to the Laplace operator on the sphere  
\begin{equation}\label{SHcasimir}
\Box_y Y^I(y)=-\frac{l(l+d')}{R_2^2} Y^I(y)
\end{equation}
and satisfy the relation
\begin{equation}\label{SHcomplete}
\sum_{\scriptscriptstyle m_1\ge\dots\ge
  m_{d'-1}\ge|m_{d'}|\ge0}^lY^I(y)Y^{\ast I}(y') = \frac{(2l+d')\Gamma
  (\frac{d'}{2})}{4 \pi^{\frac{d'}{2}+1}} 
C_l^{(\frac{d'}{2})}(\cos\Theta)\col\qquad\cos\Theta=\frac{\vec Y\cdot\vec
  Y'}{R_2^2}=1-\frac{v}{2R_2^2}\pnt  
\end{equation}
The $C_l^{(\frac{d'}{2})}$ are the Gegenbauer Polynomials which can be defined
via their generating function 
\begin{equation}\label{genfunc}
\frac{1}{(1-2q\eta+q^2)^\beta}=\sum_{l=0}^\infty q^lC_l^{(\beta)}(\eta)\pnt
\end{equation} 
%\begin{equation}\label{GegenbauerasHyper}
%C_l^{(\frac{d'}{2})}(z)=\frac{\Gamma(l+d')}{\Gamma(l+1)\Gamma(d')}\hypergeometric{-l}{l+d'}{\tfrac{d'}{2}+\tfrac{1}{2}}{\tfrac{1-z}{2}}\pnt  
%\end{equation}
%For $\text{S}^3$ one finds the explicit relation
%\begin{equation}
%\sum_{\scriptscriptstyle m_1 \ge | m_2| } ^l Y^I(y) Y^{\ast I}(y') =
%\frac{l+1}{4\pi^2}\frac{1}{i\sqrt{1-z^2}}\big((z+i\sqrt{1-z^2})^{l+1}-(z-i\sqrt{1-z^1})^{l+1}\big)\Big|_{z=\frac{y\cdot 
%y'}{R_2^2}}\pnt 
%\end{equation}
%For $\text{S}^5$ one finds the explicit relation
%\begin{equation}
%\begin{aligned}
%\sum_{\scriptscriptstyle m_1 \ge m_2 \ge m_3 \ge | m_4| }^l Y^I(y) Y^{\ast I}(y')
%&=\frac{l+2}{8\pi^3}\Big[\frac{z}{i\sqrt{1-z^2}^3}\big((z+i\sqrt{1-z^2})^{l+2}-(z-i\sqrt{1-z^2})^{l+2}\big)\\ 
%&\phantom{=\frac{l+2}{8\pi^3}\Big[}-\frac{l+2}{1-z^2}\big((z+i\sqrt{1-z^2})^{l+2}+(z-i\sqrt{1-z^2})^{l+2}\big)\Big]\Big|_{z=\frac{y\cdot
%y'}{R_2^2}}\pnt
%\end{aligned}
%\end{equation}
Using  \eqref{AdSprop} and \eqref{SHcomplete} for $\Delta_+=l+\frac{d+d'}{2}$
(leading to \eqref{modesumprop}), one finds the solution \eqref{AdSSprop} if
the following relation holds for $\alpha\ge\beta>0$, 
$2\alpha,2\beta\in\mathds{N}$
\begin{equation}\label{hyperid}
\sum_{l=0}^{\infty}\frac{\Gamma(l+\alpha)}{\Gamma(l+\beta)}\Big(\frac{\xi}{2}\Big)^l\hypergeometric{\tfrac{l}{2}+\tfrac{\alpha}{2}}{\tfrac{l}{2}+\tfrac{\alpha}{2}+\tfrac{1}{2}}{l+\beta+1}{\xi^2}C_l^{(\beta)}(\eta) 
=\frac{\Gamma(\alpha)}{\Gamma(\beta)}\frac{1}{(1-\xi\eta)^\alpha}\col
\end{equation} 
with the interpretation $\alpha=\frac{d+d'}{2}$, $\beta=\frac{d'}{2}$.
We could not find the above formula in the literature. It is in fact a
summation rule for a product of a special hypergeometric function for which so
called quadratic transformation formulae exist and which can be expressed in
terms of a Legendre function \cite{Abramowitz:1972} and of a Gegenbauer
polynomial. The identity can therefore be re-expressed in the following way 
\begin{equation}\label{polyid}
\Big(\frac{2}{\xi}\Big)^\beta(1-\xi^2)^\frac{\beta-\alpha}{2}\sum_{l=0}^{\infty}\Gamma(l+\alpha)(l+\beta)P_{\alpha-\beta-1}^{-l-\beta}\big(\tfrac{1}{\sqrt{1-\xi^2}}\big)C_l^{(\beta)}(\eta)=\frac{\Gamma(\alpha)}{\Gamma(\beta)}\frac{1}{(1-\xi\eta)^\alpha}\pnt 
\end{equation}
The simplest way to prove\footnote{We thank Danilo Diaz for delivering a
  simplification of our proof in the previous version, that allows for an
  extension to more general values of $\alpha$ and $\beta$.}
is to use the orthogonality of the Gegenbauer polynomials
\begin{equation}
\int_{-1}^{1}d\eta~
(1-\eta^2)^{\beta-\frac{1}{2}}C_m^{(\beta)}(\eta)C_n^{(\beta)}(\eta)=\frac{2^{1-2\beta}\pi\Gamma(n+2\beta)}{\Gamma(n+1)(n+\beta)\Gamma(\beta)^2}\delta_{mn} 
\end{equation}
to project out a term with fixed $l$ from the sum in \eqref{hyperid}. The 
Gegenbauer polynomials in the integral on the r.h.s.\ of \eqref{hyperid} should
then be expressed via Rodrigues' formula
\begin{equation}
C_n^{(\beta)}(\eta)=(-1)^n2^{-n}\frac{\Gamma(\beta+\frac{1}{2})\Gamma(n+2\beta)}{\Gamma(n+1)\Gamma(2\beta)\Gamma(n+\beta+1)}(1-\eta^2)^{\frac{1}{2}-\beta}\frac{d^n}{d\eta^n}(1-\eta^2)^{n+\beta-\frac{1}{2}}\pnt 
\end{equation}
Repeated partial integration to shift all the above derivatives to the first 
factor under the integral and a suitable variable
transformation at the end leads to an integral that can be expressed via a
hypergeometric function. This hypergeometric function is than connected to the
one on the left hand side of \eqref{hyperid} via the quadratic transformation
formula
\begin{equation}
\hypergeometric{a}{a+\tfrac{1}{2}}{c}{z^2}=(1+z)^{-2a}\hypergeometric{2a}{c-\tfrac{1}{2}}{2c-1}{\tfrac{2z}{1+z}}\pnt 
\end{equation}
Both sides of \eqref{hyperid} than match and the proof is complete. The
relation \eqref{hyperid} is therefore valid not only for $\alpha\ge\beta>0$,
$2\alpha,2\beta\in\mathds{N}$ but for all $\beta>0$.

\section{The chordal distance in the plane wave}
\label{chordaldistpw}
It has been shown by Penrose \cite{Penrose:1965} that it is impossible to
globally embed the plane wave spacetimes into a pseudo-Euclidean spacetime. 
However, an isometric
embedding of the $D$-dimensional CW spaces with metric
\begin{equation}\label{CWmetric}
\de s^2=-4\de z^+\de z^-+H_{ij}z^iz^j(\de z^+)^2+\de \vec
z^2\pnt
\end{equation}
in $\mathds{R}^{2,D}$ is possible \cite{Blau:2002mw}.
The flat metric of $\mathds{R}^{2,D}$ via the 
coordinate transformations
\begin{equation}
Z_+^1=\frac{1}{2}(Z_0+Z_d)\col\quad Z_-^1=\frac{1}{2}(Z_0-Z_d)\col\quad Z_+^2=\frac{1}{2}(Z_{d+1}+Z_{d-1})\col\quad Z_-^2=\frac{1}{2}(Z_{d+1}-Z_{d-1})
\end{equation}
can be transformed to    
 \begin{equation}
\de s^2=-4\sum_{k=1}^2\de Z_+^k\de Z_-^k+\sum_{i=1}^D\de Z_i^2\pnt
\end{equation}
If the hypersurface is defined as
\begin{equation}
\sum_{k=1}^2Z_+^kZ_+^k=1\col\qquad H_{ij}Z^iZ^j+4\sum_{k=1}^2Z_+^kZ_-^k=0
\end{equation}
and parameterized as follows
\begin{equation}
\begin{aligned}
Z_+^1&=\cos z^+\col\\
Z_-^1&=-z^-\sin z^+-\frac{1}{4}H_{ij}z^iz^j\cos z^+\col\\
Z_+^2&=\sin z^+\col\\
Z_-^2&=z^-\cos z^+-\frac{1}{4}H_{ij}z^iz^j\sin z^+\col\\
Z_i&=z_i\col
\end{aligned}
\end{equation}
one finds that the induced metric is given by \eqref{CWmetric}.
The chordal distance in the plane wave reads
\begin{equation}\label{pwchordaldist}
\begin{aligned}
\Phi(z,z')&=-4\sum_{k=1}^2(Z_+^k(z)-Z_+^k(z'))(Z_-^k(z)-Z_-^k(z'))+\sum_{i=1}^D(Z_i(z)-Z_i(z'))^2\\
&=-4(z^--z'^-)\sin(z^+-z'^+)+2H_{ij}(z^iz^j+z'^iz'^j)\sin^2\frac{z^+-z'^+}{2}+(\vec
z-\vec z')^2\pnt
\end{aligned}
\end{equation}
In our case where $H_{ij}=-\delta_{ij}$, this result matches 
\eqref{totalchordalpw}. 

%\section{Relation between the chordal and the geodesic distance}
%\label{relchordalgeodist}
%%%%%%%%%%%%%%%%%%%%%%%%%%%%%%
%%%%%%%%%%%%%%%%%%%%%%%%%%%%%%

\bibliographystyle{utphys}
\bibliography{references}

\end{document}